\documentclass[conference]{IEEEtran}
\IEEEoverridecommandlockouts
\usepackage{cite}
\usepackage{ulem}
\usepackage{hyperref}
\usepackage{amsmath,amssymb,amsfonts}
\usepackage{algorithmic}
\usepackage{stfloats}
\usepackage{graphicx}
\usepackage{textcomp}
\usepackage{xcolor}
\usepackage{threeparttable}
\def\BibTeX{{\rm B\kern-.05em{\sc i\kern-.025em b}\kern-.08em
    T\kern-.1667em\lower.7ex\hbox{E}\kern-.125emX}}
\begin{document}
\newcommand{\rev}[1]{{\color{blue}#1}}

\title{Causal Intervention for Measuring Confidence in Drug-Target Interaction Prediction}

\author{
\IEEEauthorblockN{Wenting Ye$^a$, Chen Li$^b$, Yang Xie$^a$, Wen Zhang$^{a,d,e}$, Hong-Yu Zhang$^{a,d,e}$, Bowen Wang$^a$, 
\\Debo Cheng$^{c*}$, Zaiwen Feng$^{a,d,e,f,g*}$}
\IEEEauthorblockA{$^a$ College of Informatics, Huazhong Agric. University, Wuhan, China}
\IEEEauthorblockA{$^b$ Graduate School of Informatics, Nagoya University, Chikusa, Nagoya, 464-8602, Japan}
\IEEEauthorblockA{$^c$ UniSA STEM, University of South Australia, Adelaide, 5095, Australia}
\IEEEauthorblockA{$^d$ Hubei Key Laboratory of Agricultural Bioinformatics, Huazhong Agricultural University, Wuhan 430070, China}
\IEEEauthorblockA{$^e$ Key Laboratory of Smart Farming for Agricultural Animals, Huazhong Agricultural University, Wuhan 430070, China}
\IEEEauthorblockA{$^f$ Hubei Hongshan Laboratory, Huazhong Agricultural University, Wuhan 430070, China}
\IEEEauthorblockA{$^g$ National Key Laboratory of Crop Genetic Improvement, Huazhong Agricultural University, Wuhan 430070, China}

}
\maketitle
\let\thefootnote\relax\footnote{*Correspondence: Debo Cheng (debo.cheng@unisa.edu.au) and Zaiwen Feng (Zaiwen.Feng@mail.hzau.edu.cn)}

\begin{abstract}
Identifying and discovering drug-target interactions (DTIs) are vital steps in drug discovery and development. They play a crucial role in assisting scientists in finding new drugs and accelerating the drug development process.
Recently, knowledge graph and knowledge graph embedding (KGE) models have made rapid advancements and demonstrated impressive performance in drug discovery. 
 However, such models lack authenticity and accuracy in drug target identification, leading to an increased misjudgment rate and reduced drug development efficiency. To address these issues, we focus on the problem of drug-target interactions, with knowledge mapping as the core technology. 
 Specifically, a causal intervention-based confidence measure is employed to assess the triplet score to improve the accuracy of the drug-target interaction prediction model. 
Experimental results demonstrate that the developed confidence measurement method based on causal intervention can significantly enhance the accuracy of DTI link prediction, particularly for high-precision models. The predicted results are more valuable in guiding the design and development of subsequent drug development experiments, thereby significantly improving the efficiency of drug development.
\end{abstract}


\section{Introduction}
A drug target is a biomacromolecule or biomolecular structure in the body, usually including a protein or nucleic acid,  intrinsically associated with a specific disease process. Drugs function by binding to these specific targets and modifying the gene function of the target to achieve the desired therapeutic effect \cite{triggle2006comprehensive}, \cite{wang2020novel}. Computational prediction of DTI is critical to advancing our understanding of drug mechanisms of action, disease pathology, and side effects\cite{ezzat2019computational}. The traditional drug development process takes 12-15 years, and costs more than \$1 billion. However, with the advancement of computer technology and bioinformatics, the identification of drug targets has increasingly turned to information technology.
Sequence and structure alignments\cite{zhang2021mcdb,ahmadi2022active}, homology modelling \cite{hegazy2022identification}, and deep learning-based approaches \cite{ozturk2018deepdta}, which have been widely used in drug-target research.

Recently, many models have emerged in the field of drug-target interactions with significant breakthroughs and effects. However, most previous studies have problems with existing confidence measurement methods, which are usually based on the original confidence score or score ranking of the model output, and the stability cannot be guaranteed. Additionally, the output scores of such models are often uninterpretable.
To solve this problem, we apply the confidence measure method of causal intervention to drug-target link prediction. The method first applies the idea of causal intervention to change the embedding representation. Then, the KGE model was used to complete a new triplet composed of drug intervention entities to score again. Finally, a new confidence score was obtained by consistency calculation. Unlike the previous approach, which only focused on the optimal score, this approach measures the robustness of the embedded representation results by actively intervening in the input of the entity vector. Our motivation is to more truly reflect the confidence probability of the model's predicted results and improve the performance of the KGE model in drug-target interaction tasks.

The implementation of domain intervention consistency \cite{wang2021neighborhood} proved that the intervention approach can improve the model confidence score, but it has not been tried in the field of drug development. This study aims to solve the problem that the model confidence is not accurate enough in the drug-target prediction problem dominated by the knowledge graph embedding model, so as to improve the accuracy and stability of drug-target prediction. The main contributions are summarized as follows.
\begin{itemize}
\item The embedding representation effects of different KGE models are investigated in two datasets applicable to the drug-target domain class.
\item The model performance is evaluated by comparing the prediction results before and after intervening in the neighborhood of the embedded vectors.
\item Probabilistic calibration is used to comprehensively evaluate the methods, and then to verify whether the confidence measure based on causal intervention can improve the confidence score of drug-target link prediction.
\end{itemize}

\section{Related Work}

Drug-target prediction is a very important field, and advanced techniques such as machine learning, deep learning and knowledge graphs have also made many achievements in this field. \cite{nascimento2016multiple} used multi-core learning and clustering methods for drug-target measurement modeling, which demonstrated its certain advantages in predictive performance. Mohamed et al 2019 proposed the Complex\cite{trouillon2016complex} model to transform the drug-target prediction problem into a link prediction problem on the knowledge graph. Experiments show that the Complex model achieves the best results compared with the traditional KGE model. Bonner et al 2022\cite{bonner2022review,bonner2022understanding}discussed the application and performance of different knowledge graph embedding models in drug discovery and compared their performance in drug target prediction, drug side effect prediction, and drug interaction prediction.

Although previous studies have made progress in using KGE technology for drug-target prediction, there are still some shortcomings and challenges. For example, TransH\cite{wang2014knowledge}, RotatE, DisMult and other models have achieved good performance in drug-target prediction, but lack interpretation of predicted results.  \cite{wang2021drug, wang2022predicting}also has the problem of poor model interpretation. Researchers in related fields have made some efforts. \cite{yamanishi2008prediction} Integrated information from the chemical and genomic space, \cite{zhang2017predicting} explored the chemical properties, biological pathways, phenotypic effects, and network interactions between drugs of binding drugs, \cite{yu2021predicting} learned and integrated characteristic representations of drugs and diseases at different convolution layers.

In order to solve the problem of the unexplainability of model output scores, we propose a new method to use causal intervention confidence measurement in drug-target link prediction. Instead of focusing only on the optimal score, this method measures the robustness of the embedded representation results by actively interfering with the input of the entity vector. Our motivation is to improve the performance of the KGE model in drug-target interaction tasks, making it a more reliable and interpretable predictive tool.

\section{The Proposed Causal Intervention Method}
In this section, we first overview our proposed causal intervention. Next, we describe the embedded intervention~\cite{cheng2022data}. The third part will introduce the domain score sequence acquisition, and the fourth part will discuss the causal intervention confidence calculation. Finally, we will discuss the drug-target link prediction calibration.
\subsection{Overview}\label{AA}
The core idea of the proposed causal intervention method is to intervene in the dimension value of a drug entity embedding vector, and then the robustness of the score is calculated as the criterion to determine whether the original triplet is real enough by comparing the triplet score sequence composed of the original drug entity and the triplet score sequence composed of the entity after the intervention.

\begin{figure*}
    \centering
    \includegraphics[width=0.92\textwidth, height=0.285\textwidth]{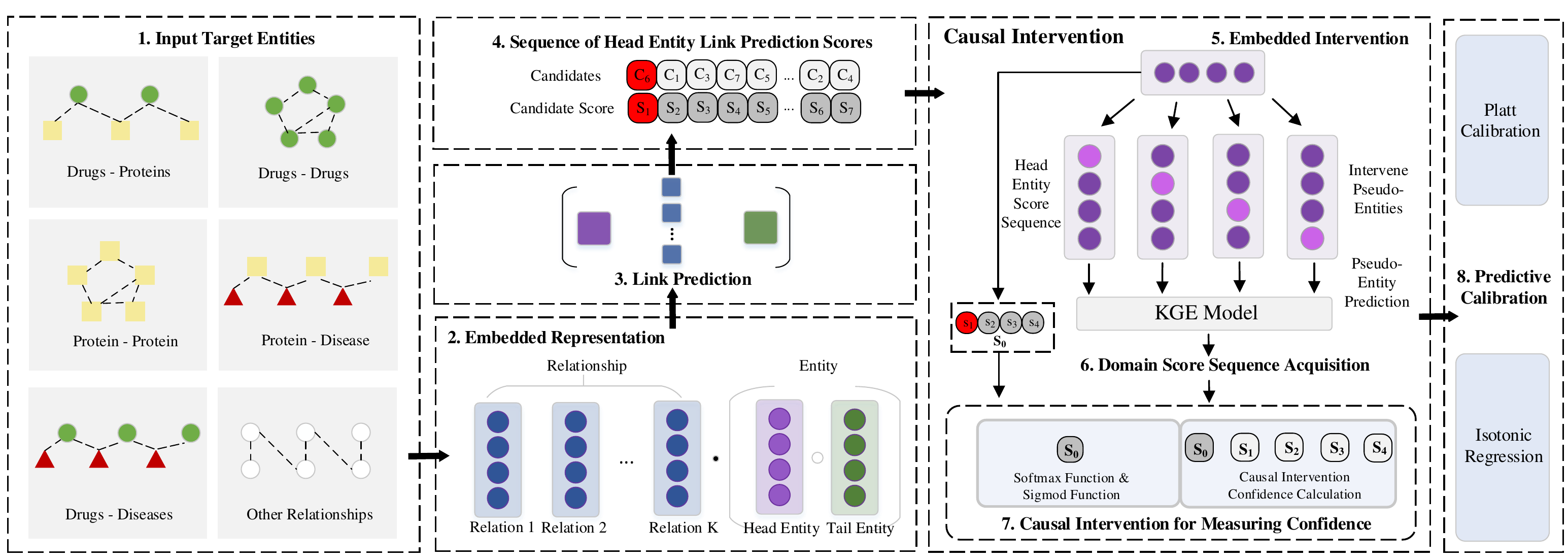}
    \caption{The schematic diagram of the proposed causal intervention method for measuring confidence in drug-target causal intervention prediction.}
    \label{fig:pipline}
\end{figure*}

We visualize the overview of our proposed causal intervention method in Fig.~\ref{fig:pipline}. It describes the main steps and processes of the proposed causal intervention method for confidence measurement on drug-target interaction prediction. Firstly, target entities such as drugs and proteins in the data, along with their biological relationships, are embedded using the KGE model for learning latent representation. Subsequently, the learned representation is employed to predict the links between drugs and targets and their relationships in the dataset. For each triplet, the model generates a score, forming a score sequence. Next, the intervention entity is embedded to assess the confidence of the causal intervention. Finally, the results are calibrated using two common probabilistic calibration methods. The causal intervention method is then compared with the confidence method in traditional link prediction tasks to verify the effectiveness of the proposed causal intervention method in drug-target link prediction. In the following sections, we provide a description of the detailed steps of the proposed causal intervention method for the confidence measurement method.

\subsection{Embedded Intervention}
In this paper, our main focus is on studying the impact of causal intervention on the representation characteristics of embedded vectors. By intervening in the embedded vectors, we can manipulate the representation eigenvalues to observe how such changes affect the output results of the model~\cite{pearl2009causality}. For a drug-target entity represented by a $D$-dimensional embedding vector with $d$ features, we explore different intervention strategies to interfere with the eigenvalues, resulting in $d$ different representation vectors. The only difference between these $d$ vectors and the initial drug-target entity's embedding representation lies in the variation of each dimension value. However, this change may cause some loss of the original semantic information. To assess the accuracy and robustness of the semantic representation, we compare the influence of the vector representation after causal intervention with the original vector representation on the link prediction results. The core idea of the proposed causal intervention method is to determine whether the original entity embedding representation possesses sufficient anti-interference and representativeness, analyzed from the perspective of stability measurement.


The current design of the intervention function has several limitations. One such limitation is its tendency to intervene more on low-dimensional vectors, as high-dimensional data vectors are typically denser, making intervention and interpretation more challenging.  In practical applications, dimension reduction is often performed on embedded vectors to enhance their feature representation interpretability. Furthermore, careful consideration should be given to the selection of the number of intervention dimensions. If the number is too large, it may lead to the vector transitioning from one semantic category to another, thus adversely affecting the performance of the model and hindering the achievement of the desired intervention outcomes.


\subsection{Domain Score Sequence Acquisition}
Drug-target link prediction tasks can be divided into three categories: head entity prediction, relationship prediction, and tail entity prediction. In this paper, the tail entity is adopted as the prediction method for the link prediction task. Specifically, each triplet $<h, r, t>$ is modified by replacing the tail entity $t$, and the trained KGE model is used to compute a new score for each prediction triplet. The scores obtained for each replacement tail entity form a set of original score sequences, denoted as $S$. For the KGE model, the optimal score does not necessarily imply that the triplet is the most realistic triplet. Therefore, in confidence scoring, unlike the traditional focus on the optimal score, this work evaluates the confidence score by predicting the robustness of the embedded entity.

\begin{figure}[t]
\centering
\includegraphics[width=0.76\linewidth, height=0.118
\textwidth]{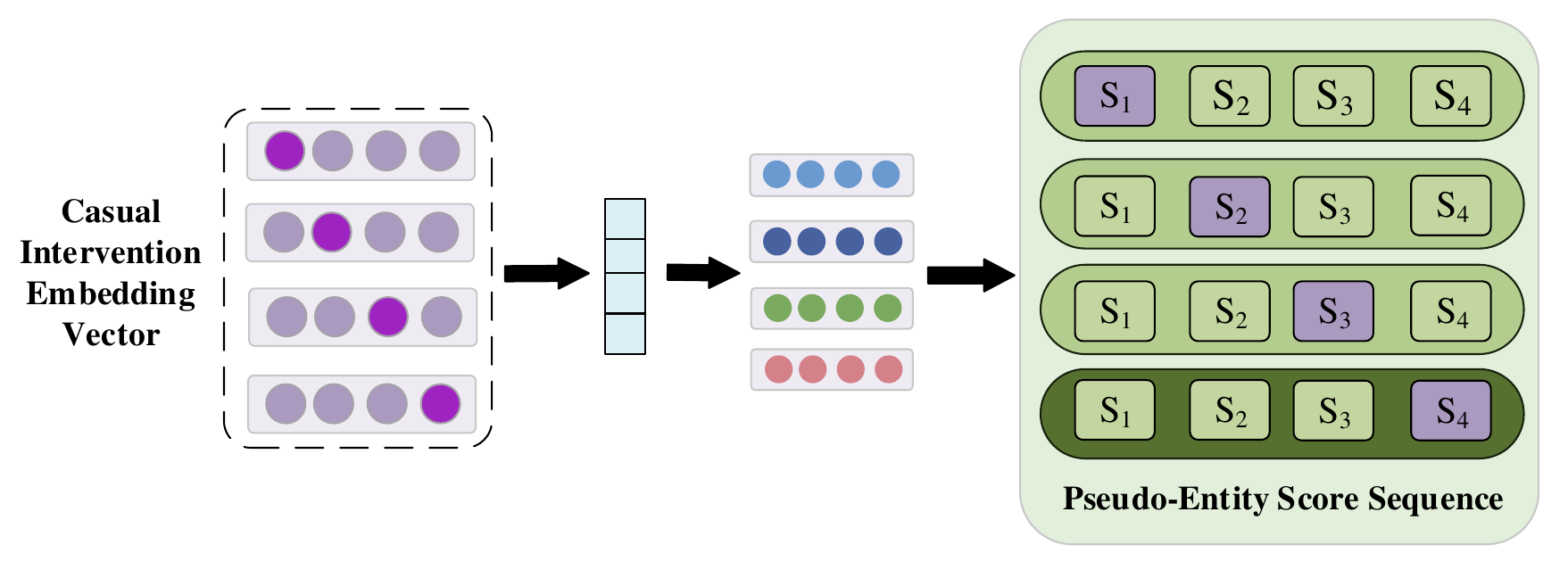}
\caption{Schematic diagram of pseudo entity score sequence}
\label{fig:scoreSequence}
\end{figure}

As shown in Fig.~\ref{fig:scoreSequence}, the drug entity undergoes a causal intervention operation, leading to the generation of a group of domain intervention entities by altering the embedded representation values in each dimension. During the link prediction process, the KGE model scores each triplet formed by the domain pseudo-entities and the corresponding relationships between the original entity and each target prediction entity. Suppose a 4-dimensional embedded entity $e$, a relation $r$, and four candidate target tail entities $T = \left\{t_1, t_2, t_3, t_4\right\}$ are predicted by the $e-r$ query. After the intervention, four domain pseudo-entities $E^{'} = \left\{e_1^{'}, e_2^{'}, e_3^{'}, e_4^{'}\right\}$ are generated. The model then scores the prediction triplet formed by each pseudo-entity $e^{'}$, relation $r$, and four tail entities $T$, generating the score sequence $S^{'} = \left\{S_1^{'}, S_2^{'}, S_3^{'}, S_4^{'}\right\}$ corresponding to each pseudo-entity $e^{'}$. As a result, there are $d$ score sequences, and the total number of scores is $d \times m$, represented as $S_{i}$. Next, the scoring sequence $S_{i}^{'}$ corresponding to each pseudo-entity $e_{i}^{'}$ and the scoring sequence $S$ corresponding to the initial entity $e$ are used to calculate the consistency difference, ultimately yielding a new consistency score.

\subsection{Causal Intervention for Measuring Confidence}
\begin{figure}[t]
\centering
\includegraphics[width=0.76\linewidth, height=0.18\textwidth]{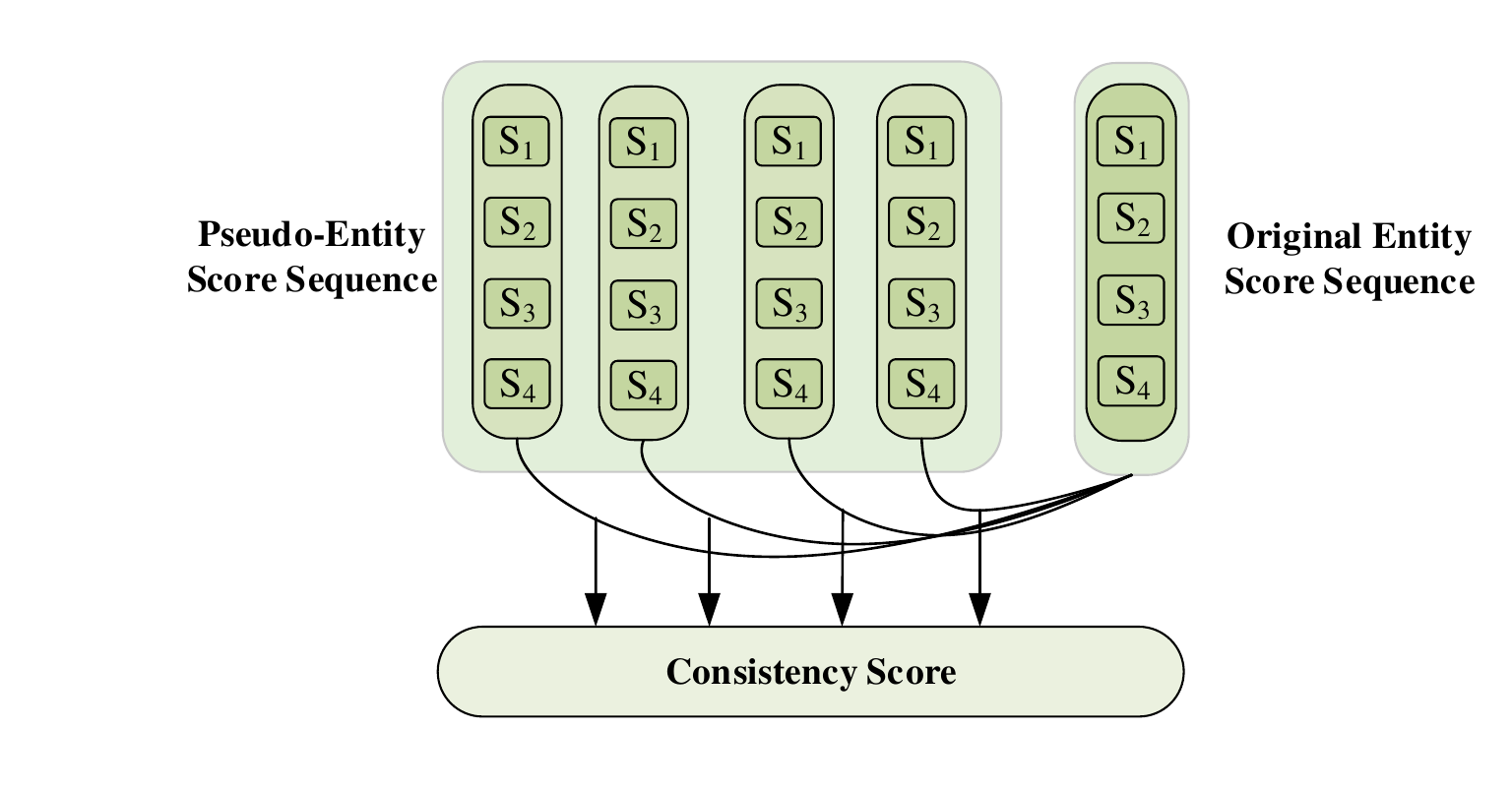}
\caption{Schematic diagram of consistency score calculation}
\label{fig:consistency_score}
\vspace{-15pt}
\end{figure}

The confidence score, also known as the consistency score, measured by the causal intervention method, indicates the stability of an entity's embedding. As depicted in Fig.~\ref{fig:consistency_score}, the primary calculation involves assessing the consistency of each field's score sequence in the pseudo-entity score sequence with the original entity score sequence, followed by normalization of the consistency values. The underlying assumption is that when an entity is true in the triplet of link prediction, its embedded characteristics align with the model's score prediction. Additionally, if a dimension feature in the entity embedding vector is altered, its effect on the entity is relatively small. Consequently, comparing the consistency of the two scoring sequences becomes crucial.

For the domain sequence, the index ordering of the first $K$ candidate entities is denoted as $\rho(S^{N_i})$, and the index robustness scores of the first $K$ candidate entities in the initial score sequence are calculated as follows: 

\begin{scriptsize}
\begin{equation}
\label{eq:e1}
\begin{aligned}
\text{Consis}(S, S^{N_i}) &= \sum_{j=0}^J \text{Softmax}(S)_j \cdot
(1 - \text{Sgn}(\left|\rho(S)_j - \rho(S^{N_i})_j\right|)),
\end{aligned}
\end{equation}
\end{scriptsize}

 \noindent where $J$ is a hyperparameter acting as a threshold to select the first $J$ indexes that fit. A robustness score value of $1$ indicates that the index of the domain score sequence matches that of the initial sequence. A lower robustness score implies a more mismatched index ranking based on the first $J$ positions, indicating a larger degree of difference and worse robustness. After the intervention of a feature, the robustness score is obtained as $\text{Consis}(S, S^{N_i})$. Then, for the entity embedding vector with dimension $d$, the normalized confidence score can be obtained as follows:

\begin{small}
\begin{align}\label{eq:e2}
    \text{P}_{ci}(S) = \frac{1}{d}\sum_{i=0}^d\text{Consis}(S, S^{N_i}).
\end{align}
\end{small}

\subsection{Drug-target Link Prediction Calibration}
In the link prediction task, model calibration is of great importance. Because the machine learning model may produce uncertain prediction results, leading to overestimated or underestimated probability values. This reduction in the reliability of predictions increases the cost of research and development and raises the risk of misdiagnosis. Therefore, proper model calibration is crucial to improve the accuracy and dependability of the predicted outcomes.

The specific operation step of calibration involves normalizing the confidence score of a prediction result to the interval [0, 1], and dividing this interval into $n$ bin. For each bin, the difference between the average confidence score and the average true score for all of the pretests is calculated. A weighted average of all bin scores is then computed to obtain an ECE (Expected Calibration Error) score\cite{niculescu2005predicting}.

In a word, the purpose of model calibration is to increase the reliability of model prediction results, allowing it to better reflect the actual prediction outcomes of samples. A smaller ECE score indicates that the confidence score of the model is closer to the real score, implying higher reliability. In drug-target link prediction, the introduction of the causal intervention method aims to more accurately evaluate the confidence score of the model and reduce the error of probabilistic calibration. This holds significant importance for drug research and development, as it helps enhance the accuracy and trustworthiness of the model's predictions.

\begin{small}
\begin{align}\label{eq:e3}
    \text{ECE}=\sum_{b=1}^B\frac{n_b}{N}\left|acc(b)-conf(b)\right|,
\end{align}
\end{small}

\noindent where  $b$ represents the number of samples in this bin, $acc(b)$ represents the true and accurate average value of samples in this bin, and $conf(b)$ represents the average predicted value of samples in this bin.

\section{Experiments}
\subsection{Experimental Setup}
Two public knowledge graphs applicable to the field of drug-target discovery, Hetionet and BioKG, were used in this experiment. Additionally, six KGE models were trained according to the original setting.
The dimension of the high-dimensional model is set to 200. For the low-dimensional model, the embedded dimension chosen in this paper is 50. 
For the hyperparameters of causal intervention, we use the entity of the former TopK with parameters {3, 5, 10, 100, 200, 300}, and for the first $J$ relative positions in the consistency calculation, the parameters are {3, 5, 10, 100}.
All models are implemented based on the Pykeen framework \cite{ali2021pykeen}, and all algorithms are implemented using the PyTorch framework based on Python. All experiments were run on NVIDIA RTX A400 GPU and Intel(R) Xeon(R) Silver 4210R CPU @ 2.40GHz on Centos.

\subsection{Evaluation Measures}
In order to fully reflect the effect of the link prediction task, we introduce three evaluation indicators:
\begin{itemize}
\item \textbf{ECE}: Expected Calibration Error is commonly utilized to evaluate the calibration effect, the lower the ECE, the better the calibration effect of the model.
\item \textbf{ACC}: Triple accuracy refers to the average accuracy of triples in the same confidence level.
\item \textbf{T10ACC}: the average accuracy of the top 10$\%$ high-confidence triples. The higher the T10ACC, the better the confidence measurement effect of the confidence measurement method. 
\end{itemize}

\begin{table}[!t]
\caption{ECE of different confidence measures on drug-target interaction prediction models}
\centering
\resizebox{.3\textwidth}{!}{
    \begin{threeparttable}
    \begin{tabular}{|c|c|c|c|c|c|c|}
        \hline
        ~ & \multicolumn{3}{|c|}{Hetionet}  &  \multicolumn{3}{|c|}{BioKG} \\
        \cline{2-7} 
       model & \multicolumn{3}{|c|}{ECE $\downarrow$ } & \multicolumn{3}{|c|}{ECE $\downarrow$ }  \\ 
       \cline{2-7}
       ~ & SIG & TOP & CI & SIG & TOP & CI  \\
       \hline
      TransE & .058 & .032 & \textbf{.016} & .150 & .106 & \textbf{.063} \\
      \hline
      Complex & .094 & .042 & \textbf{.035} & .154 & .088 & \textbf{.027} \\
      \hline
      DisMult & .051 & .037 & \textbf{.024} & \textbf{.026} & .062 & .041 \\
      \hline
      RotatE & .035 & .022 & \textbf{.020} & .083 & .093 & \textbf{.063} \\
      \hline
      TransH & .081 & .027 & \textbf{.013} & .034 & .035 & \textbf{.027} \\
      \hline
      CrossE & .067 & .037 & \textbf{.021} & .018 & .031 & \textbf{.016} \\
    \hline
    \end{tabular}
    \label{tab1}
    \begin{tablenotes}
    \footnotesize
    \item[$\star$]The values in bold indicate optimal (minimum) values.
    \end{tablenotes}
    \end{threeparttable}
}
\end{table}

\begin{table}[!t]
\caption{T10ACC of different confidence measurement methods for drug-target interactions}
\centering
\resizebox{.405\textwidth}{!}{
    \begin{threeparttable}
    \begin{tabular}{|c|c|c|c|c|c|c|c|c|}
        \hline
        ~ & \multicolumn{4}{|c|}{Hetionet}  &  \multicolumn{4}{|c|}{BioKG} \\
        \cline{2-9} 
       model & \multicolumn{3}{|c|}{T10ACC $\uparrow$} & ACC $\uparrow$ & \multicolumn{3}{|c|}{T10ACC $\uparrow$} & ACC $\uparrow$ \\ 
       \cline{2-9}
       ~ & SIG & TOP & CI & H10 & SIG & TOP & CI & H10  \\
       \hline
      TransE & .146 & \textbf{.237} & .121 & .117 & .453 & .398 & \textbf{.692} & .141\\
      \hline
      Complex & .022 & .027 & \textbf{.132} & .061 & .015 & .036 & \textbf{.064} & .013 \\
      \hline
      DisMult & .419 & .301 & \textbf{.428} & .168 & .195 & \textbf{.322} & .232 & .053 \\
      \hline
      RotatE & .299 & .453 & \textbf{.475} & .243 & .397 & .621 & \textbf{.650} & .238 \\
      \hline
      TransH & .278 & .139 & \textbf{.380} & .192 & .455 & .407 & \textbf{.588} & .183 \\
      \hline
      CrossE & .442 & .249 & \textbf{.521} & .154 & .158 & .227 & \textbf{.394} & .066 \\
    \hline
    \end{tabular}
    \label{tab2}
    \begin{tablenotes}
    \footnotesize
    \item[$\star$] The values in bold indicate optimal (maximum) values.
    \end{tablenotes}
    \end{threeparttable}
}
\end{table}

\subsection{Comparative Analysis of ECE and T10ACC}
The accuracy effects of 3 confidence measures on 6 different KGE prediction models were compared, and the causal intervention confidence measures were compared with the traditional SigmodidMax(SIG) and TopKSoftmax(TOP) confidence measures, which have been applied in very recent literature\cite{tabacof2019probability, safavi2020evaluating}. ECE and T10ACC scores for each model on different data sets were calculated separately. The two typical calibration methods used in this experiment are Platt Scaling(Platt)\cite{platt1999probabilistic} and Isotonic Regression(Isoto)\cite{guo2017calibration}.   


Table \ref{tab1} shows the ECE of different confidence measures for drug-target interaction prediction models. For ECE indicators, the SIG is generally higher than the TOP method in the Hetionet data set, and the CI has more obvious effects than the other two methods, and obtains lower ECE error values, indicating that the causal intervention method has achieved results in the stability of embedded representation, making the semantic representation of each entity and relationship more realistic.



For Hetionet in Table \ref{tab2}, the CrossE model improves the low-precision prediction results to a relatively high confidence accuracy of 0.521 by the CI, indicating that the accuracy of prediction results of this model is better than other baseline models. 
However, the overall confidence in the complex model was low due to low ACC scores.
In the Hetionet dataset, the ACC of Complex is higher than that in the BioKG dataset. The T10ACC scores based on the SIG and the TOP were similar, but the CI was nearly twice as high, indicating that the causal intervention method would show more obvious effects with high ACC scores, while the performance was relatively poor with low accuracy.

\subsection{Comparative Analysis of Confidence Measures}
This experiment focuses on the differences in probabilistic calibration of different confidence measurement methods for a single KGE model. In this experiment, the RotatE model was selected as the reference model, and three calibration modes, non-calibration, the Platt and the Isoto, were used to analyze the experimental results, where the $X-axis$ represents the average prediction probability and the $Y-axis$ represents the actual accuracy.

\begin{figure*}[t]
\centering
\includegraphics[width=0.8\textwidth, height=0.3\textwidth]{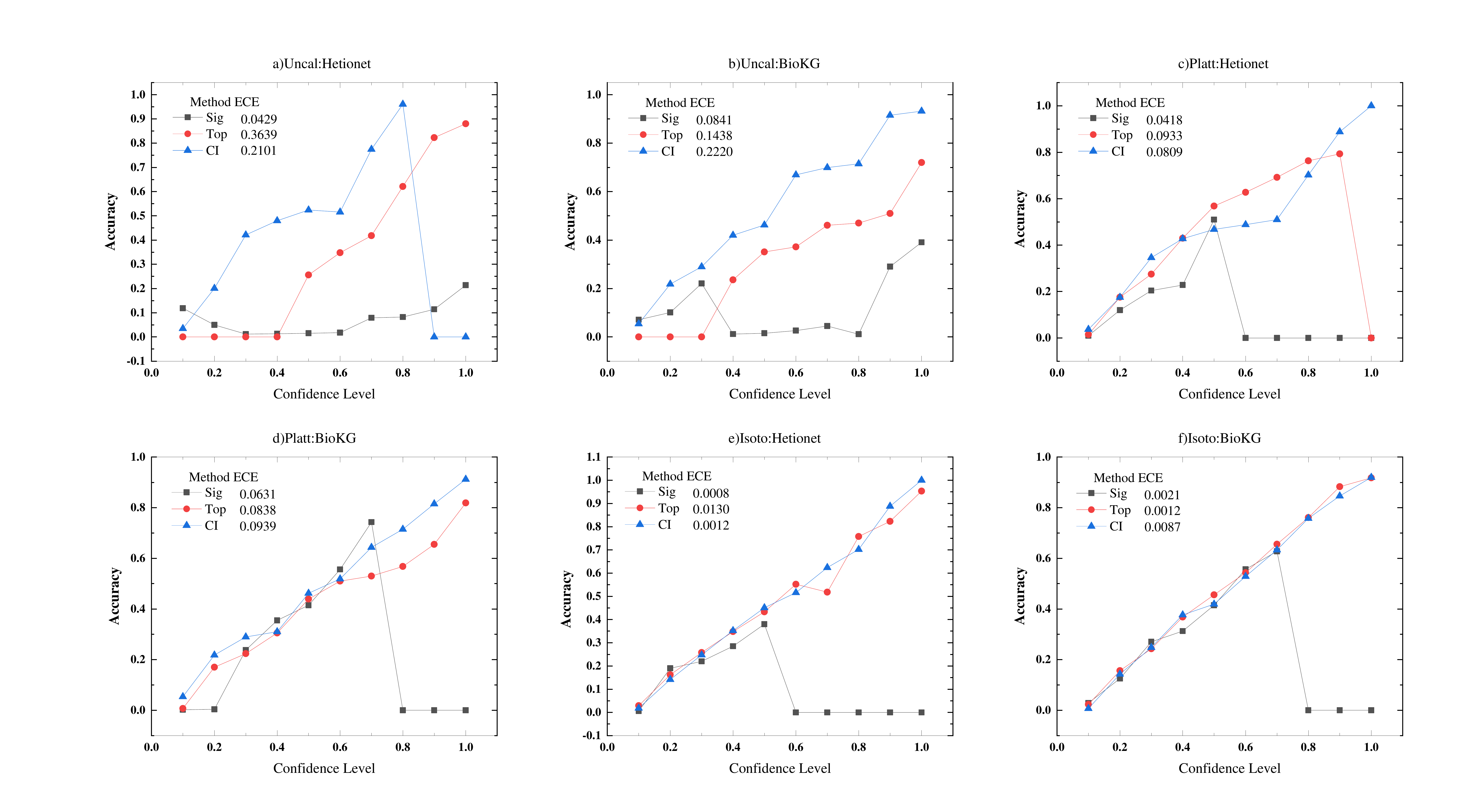}
\caption{Reliability diagrams of the RotatE model using different confidence measurement and calibration methods for two datasets: Hetionet(a) and BioKG(b).
}
\label{fig:Exp_Res}
\end{figure*}

As shown in Fig.~\ref{fig:Exp_Res}a)--b), the SIG function without calibration performs more clearly for high and low confidence scores, while the Top mainly targets high confidence scores. Compared with the SIG, the TOP is more uniform, and the CI can better predict the drug-target interactions.

The Platt for calibrating the model is shown in Fig.\ref{fig:Exp_Res}c)--d). The ECE value has a significant decline, in which the decline of the SIG and the TOP is more obvious, while that of the CI is relatively small. Compared with Fig.~\ref{fig:Exp_Res}a)--b), it can be seen that the uncalibrated model is not realistic enough to predict drug-target interactions, and the error is relatively large. Especially for the traditional confidence measurement method, it is often untrue in the high confidence score range. Using the CI, the confidence score is relatively uniform in the overall probability interval distribution, and the ECE decline is relatively low before and after calibration, which also reflects the robustness of the method based on causal intervention compared with the traditional method.

The Isoto for calibrating the model is shown in Fig.\ref{fig:Exp_Res}e)--f). The overall calibration fit of the Isoto is better, especially in the TOP and the CI, and ECE indexes are also lower. Among them, the CI is smoother than the TOP, which is close to the optimal calibration baseline. In general, the CI and the Isoto for model calibration show significant effects in both biological data sets.

\subsection{Hyperparameter Experimental Analysis}

In the causal intervention, the replacement of each target entity in the predicted triplet will generate corresponding scores. Due to the large number of tail entities, it is necessary to select the tail entity with a high TopK score, and then intervene in the embedded representation of the head entity of the triplet, and combine the TopK tail entity into a triplet to score again. The first $J$ intervention entities are selected as the number of consistent entity measures. Therefore, this experiment analyzed the influence of parameter changes on the results of causal intervention. In this experiment, the DisMult model was selected for testing on the Hetionet dataset to explore the influence of hyperparameters $K$ and $J$ on the causal intervention methods, including the values of ECE and T10ACC. Where, the selected value of $J$ is $\left\{3, 5, 10, 100\right\}$, where the value of $K$ should be greater than or equal to $J \leq K$, so the initial value of $K$ changes with the change of $J$, in this experiment $J \leq K \leq 300$.

\begin{figure}[t]
\centering
\includegraphics[width=0.7\linewidth, height=0.18\textwidth]{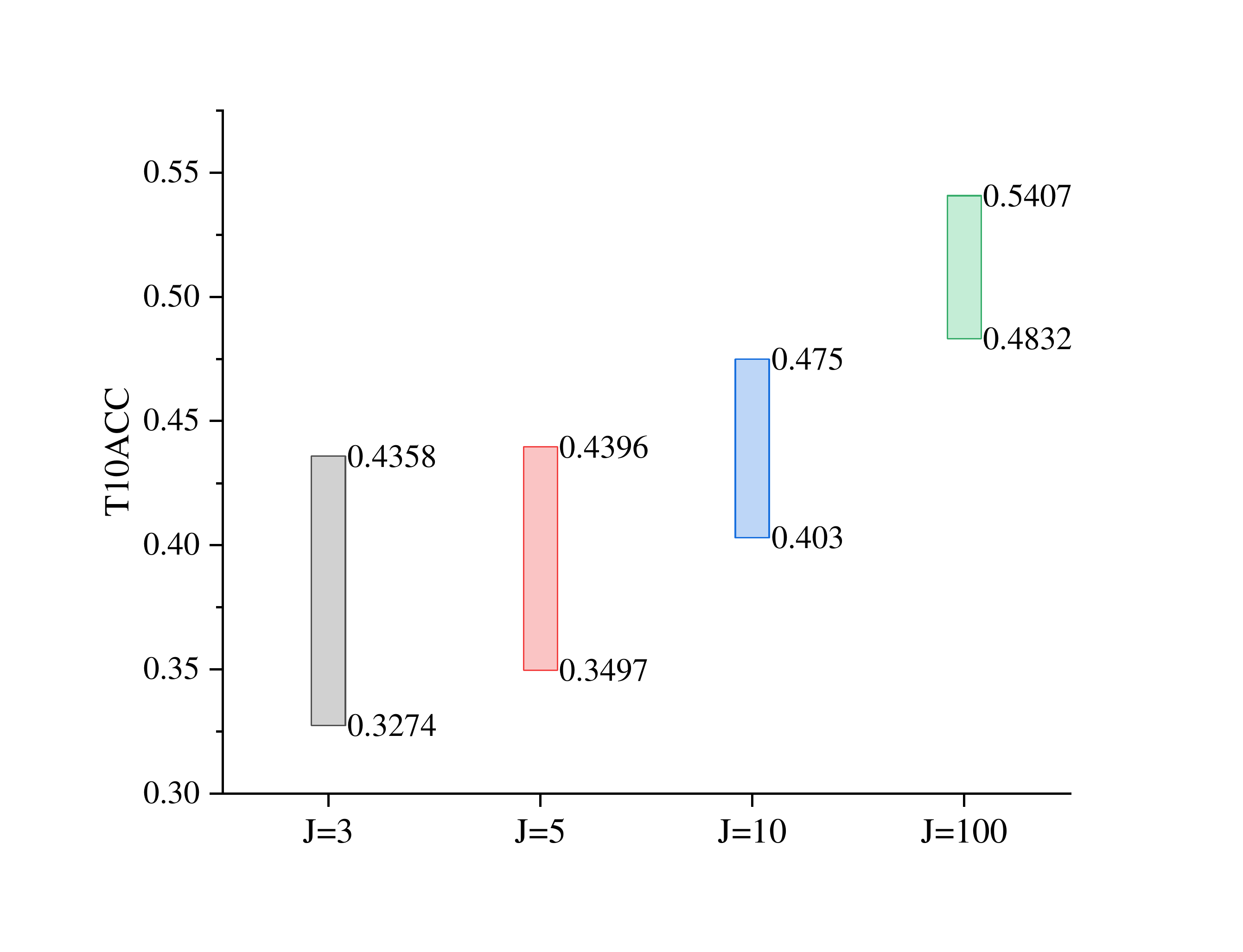}
\caption{Effect of hyperparameters K and J on T10ACC}
\label{fig:effect_kj}
\end{figure}

As shown in Fig.\ref{fig:effect_kj}, it is the interval distribution diagram of hyperparameters $K$ and $J$ for T10ACC indicators. The bottom of each column interval indicates the lowest T10ACC value, and the hyperparameter $K = J$. The value above each column interval represents the highest T10ACC value, but it does not increase with the increase of K. Due to computational power limitations, there is a significant decrease when $K$ is set to 300, which also indicates that T10ACC can only improve the accuracy of relatively high confidence triples. By comparing the choices of different $J$, it can be found that T10ACC increases with the increase of parameter $J$ and when $K = J$, which proves that there are more values in the score sequence of the selected intervention entity and the original score sequence, and the consistency comparison is more accurate.


\section{Conclusion}
In this paper, a causal intervention confidence measure was used to measure drug-target interactions prediction. Experimental results show that the proposed method improves the reliability and confidence of the model compared with the traditional method. However, it is more suitable for low-dimensional embedding representation. In the future, it is necessary to further verify the embedding choice of different dimensions and determine the appropriate dimension vector that takes into account accuracy and efficiency. All in all, this method is a beneficial supplement to the improvement of model performance in the field of drug-target, and has important significance in accelerating the process of new drug research and development, reducing research and development costs, and improving drug safety.

\section*{Acknowledgment}
This research project was supported in part by the Major Project of Hubei Hongshan Laboratory under Grant 2022HSZD031, and in part by the Fundamental Research Funds for the Chinese Central Universities under Grant 2662023XXPY004, 2662022JC004, and in part by the open funds of the National Key Laboratory of Crop Genetic Improvement under Grant ZK202203, Huzhong Agricultural University, and in part by the Inner Mongolia Key Scientific and Technological Project under Grant 2021SZD0099. Debo's research was supported in part by the Australian Research Council (under grant DP230101122).

\bibliographystyle{IEEEtran}
\normalem
\bibliography{refs}

\end{document}